\documentclass[%
 aps,
 pra,
 amsmath,amssymb,
reprint,%
]{revtex4-1}

\usepackage{graphicx}
\usepackage{bm}

\usepackage{bm}
\usepackage{graphicx}
\usepackage{color}

\newcommand{\ep}{\varepsilon}
\newcommand{\te}{\theta}
\newcommand{\tT}{{\cal T}}

\newcommand{\chiab}{\chi_{a\to b}}

\newcommand{\vA}{\bm{A}}
\newcommand{\vB}{\bm{B}}
\newcommand{\vecr}{\bm{r}}
\newcommand{\vn}{\bm{n}}

\newcommand{\xxx}{\vspace*{8pt}}

\begin{document}


\title{Persistent current in quantum torus knots}
\author{Hiroyuki Shima}%
\email{hshima@yamanashi.ac.jp}
\affiliation{%
Department of Environmental Sciences and 
Interdisciplinary Graduate School of Medicine and Engineering, 
University of Yamanashi, 4-4-37, Takeda, Kofu, Yamanashi 400-8510, Japan
}%

\date{\today}

\begin{abstract}
I consider the quantum interference of electrons
moving along knotted trajectories under external magnetic field.
The induced persistent current is formulated in terms of characteristic parameters
that classify the torus knot geometry.
The current is found to show a periodic oscillation whose period depends
strongly both on the knot's class and the field direction.
The shift in the oscillation period caused by geometric distortion
is also discussed.
\end{abstract}


\maketitle



\begin{figure}[bbb]
\centerline{
\includegraphics[width=2.7cm]{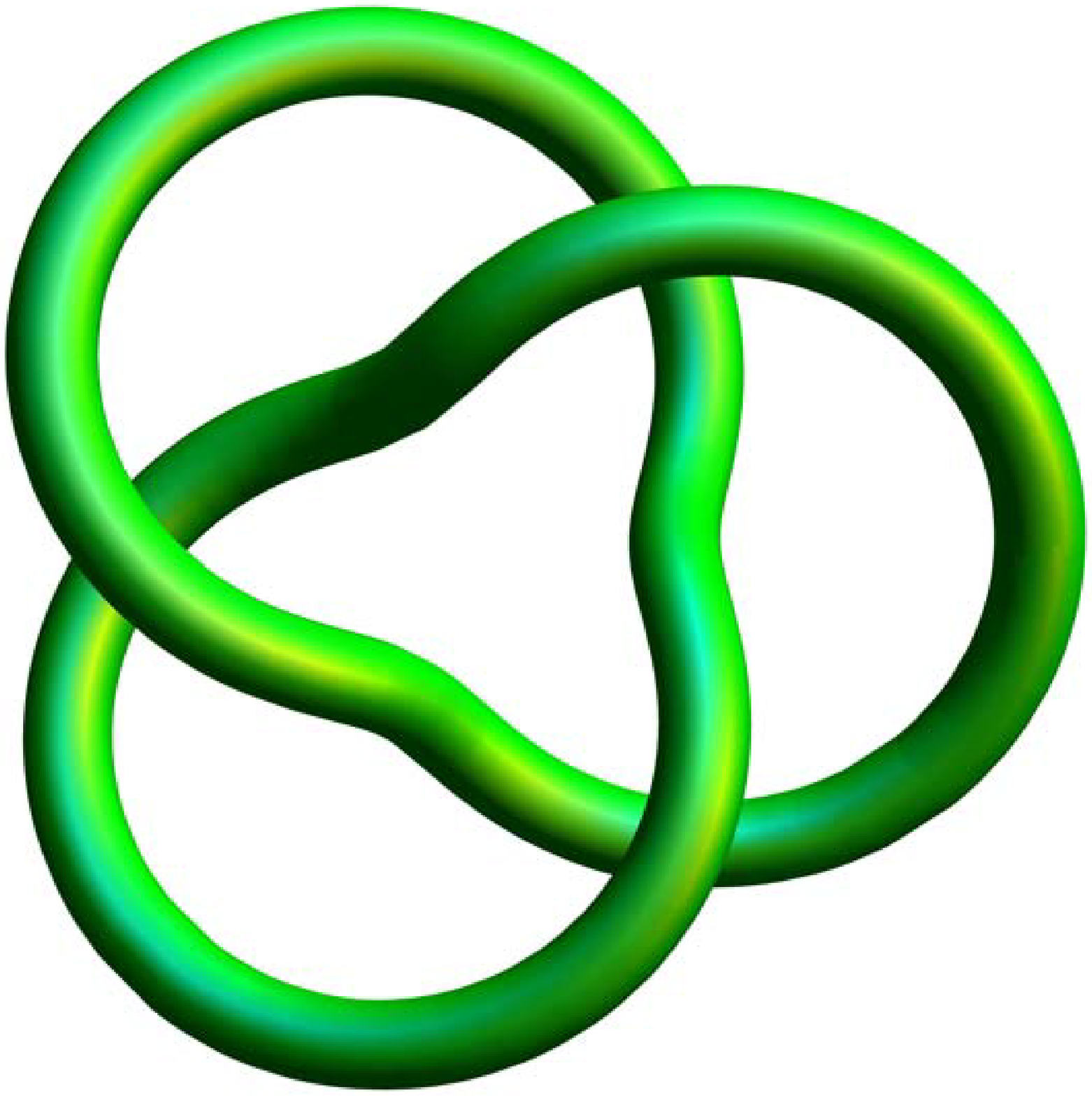}
\includegraphics[width=2.7cm]{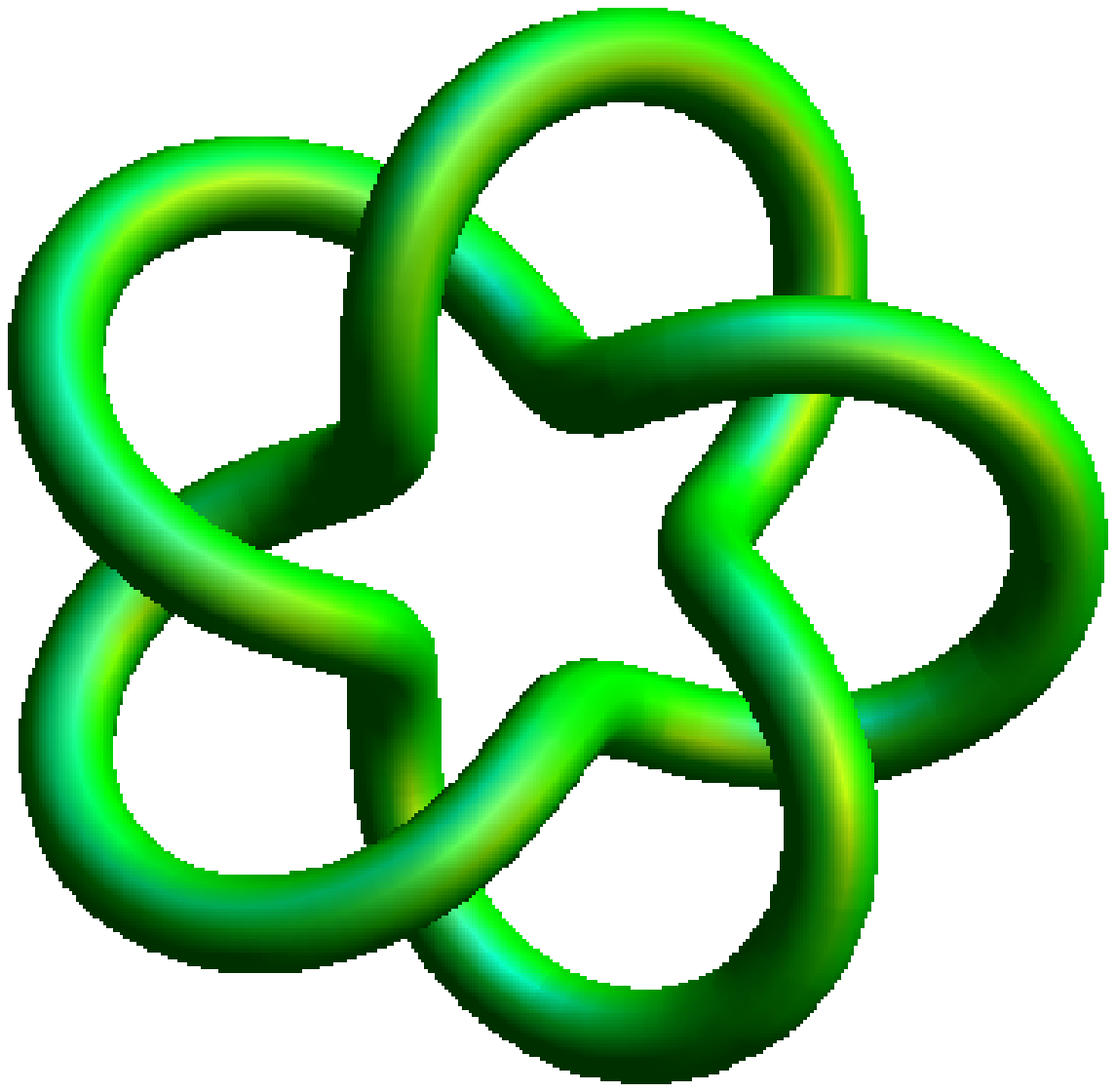}
\includegraphics[width=2.7cm]{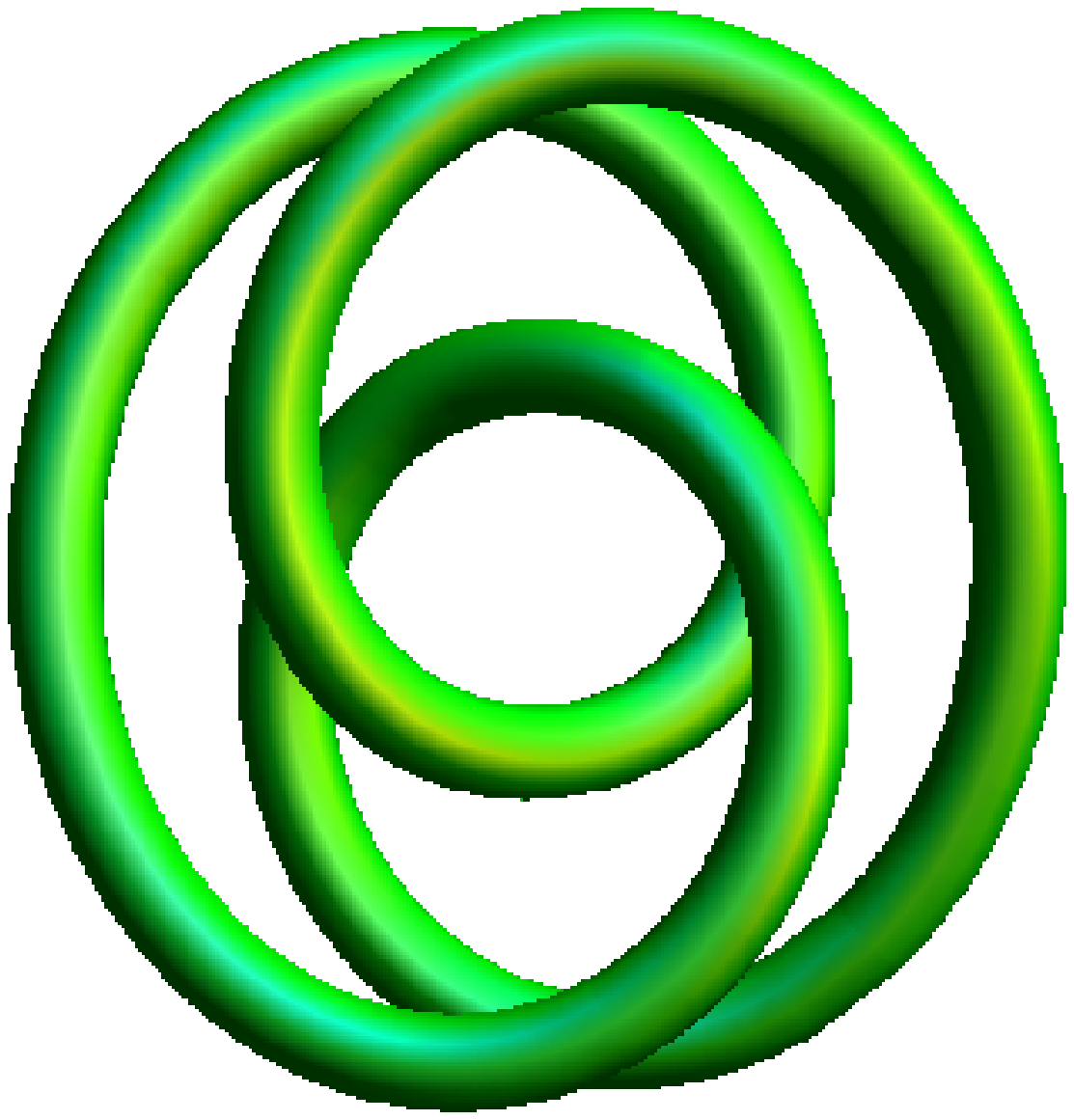}
}
\centerline{
\includegraphics[width=2.7cm]{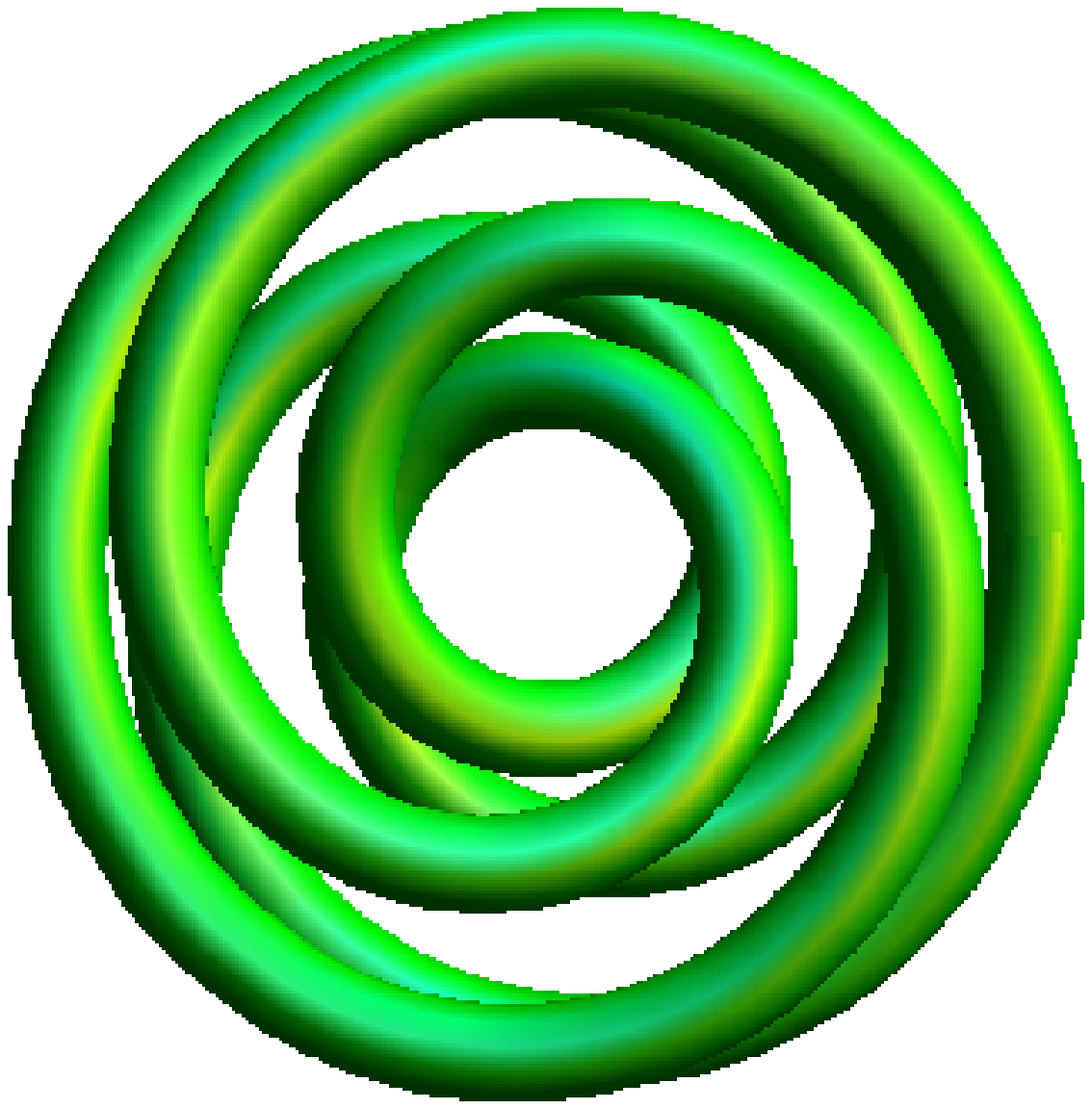}
\includegraphics[width=3.6cm]{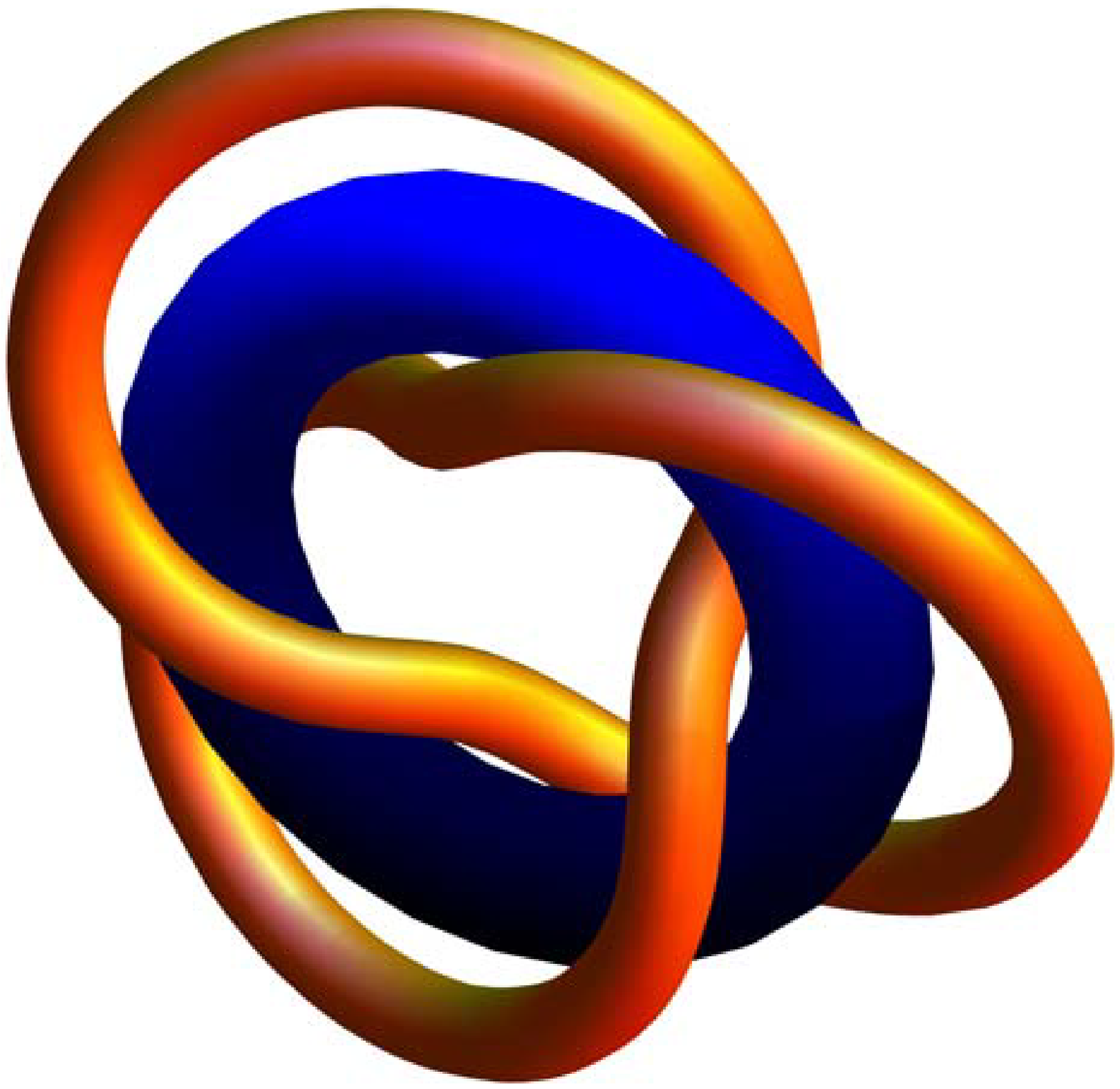}
}
\caption{(color online) Diagram of the four simplest torus knots labeled by $(p,q) = (2,3), (2,5), (3,2), (5,2)$,
respectively.
The lower right panel illustrates how the $(2,3)$ torus knot wraps around the underlying torus
colored in blue.}
\label{fig_1}
\end{figure}

State-of-the-art fabrication techniques have enabled to realize
microscopic ``knotted" objects based on organic and/or inorganic materials.
A knot means a self-entangled closed loop that cannot unraveled except by cutting the loop;
Figure~\ref{fig_1} illustrates the simplest examples of knots.
The presence of knotted molecules was first identified in DNA \cite{LFLiuJMB1976,KrasnowNature1983},
and then naturally occurring proteins \cite{CZLiangJACS1994,TakusagawaJACS1996}.
After the field of chemical topology blossomed,
diverse molecular knots endowed with different types of topology 
have become in reality \cite{ForganChemRev2011}.
Added to the molecular-scale entities,
micrometer-scale knots in carbon nanotube bundles \cite{VilatelaAdvMater2010}
and those made of triblock copolymer \cite{SchappacherAngewChemIntEd2009}
have also been realized.
To elucidate the knotted effects on their physical properties,
the isolation and precise characterization of their shapes are required.
However, the task is still a challenge even today,
which is why little has been reported for the shape-property relation
in knotted materials \cite{KimballPRL2004,BuniyPLA2008,AtanasovPLA2009}.


\xxx

An important consequence of the knotted structures,
provided they are metallic,
is thought to arise in magnetic response of internal mobile carriers.
Generally when a closed loop of a quantum wire
is threaded by a magnetic flux $\Phi$,
a persistent current will be elicited due to quantum interference
of electrons along the loop.
For instance, a micrometer-diameter ring can
support a persistent current of $I\sim$ 1 nA at temperatures $T\le$ 1 K \cite{BluhmPRL2009},
and the magnitude of the induced current
oscillates with increasing $\Phi$ with a period of the flux quantum $\Phi_0 \equiv h/e$.
This scenario may hold true no matter if the loop is knotted or unknotted.
For knotted cases, however, the flux is tangled in self-intersecting, multiple surfaces
surrounded by the self-avoiding closed loop;
hence, the resulting persistent current possibly depends on the field direction
and the knotted geometry. 
Evaluating these dependences will
facilitate the development of molecular machines and other bottom-up nanodevices
that operate under magnetic field.

\xxx

In this Brief Report, I describe the persistent current phenomena
arising in a certain group of knots, called the torus knots.
Persistent currents that occur in general
torus knots are formulated as a function of geometric parameters
of the knots,
which reveals the strong anisotropy of the current magnitude
with respect to the direction of the external magnetic field.
Analytic expression for the period of the current oscillation with increasing the field strength
is also derived,
showing the way how the knotted geometry affects the oscillation period.
These findings will provide the first step toward quantitative analyses
on the magnetic response of 
realistic molecular knots and other micrometer-scale knotted conductors.


\xxx

To proceed analytic discussion, we employ a one-dimensional coherent electron system
having a torus knot geometry.
Torus knots consist of a special group of knots, satisfying the condition that 
they all lie on a donut-shaped surface.
The class of a torus knot is uniquely identified by a pair of relatively prime
integers $p$ and $q$ \cite{Cromwell_book}; 
the one integer $p$ specifies the number of times the curve wraps around the rotational axis of the torus,
and the other $q$ specifies the number of times the curve passes through the hole of the torus.
If $p$ and $q$ are not co-prime, then we have a collection of two or more identical knots
that all lie on the same torus.
The lower panel of Fig.~1 shows
schematically the way how a (2,3) torus knot
wraps around an underlying donut-shaped surface.

\xxx

In terms of Cartesian coordinates, the $(p,q)$-torus knot is parameterized by
\begin{eqnarray}
x(\te) &=& \Big[ R + \ep \cos (q \te) \Big] \cos (p \te), \nonumber \\
\quad
y(\te) &=& \Big[ R + \ep \cos (q \te) \Big] \sin (p \te), \nonumber \\
\quad
z(\te) &=& \ep \sin (q \te) \label{eq_xyz} \label{eq_T_def}
\end{eqnarray}
with a variable $\te$ $(0\le \te \le 2\pi)$.
The two constants $R$ and $\ep$ $(0< \ep < R)$
are called by the major and minor radii of the torus, respectively;
$\ep$ is the radius of the tube that consists of the torus,
and $R$ is the radius of the circle made of the tubular axis.
It follows from Eq.~(\ref{eq_T_def}) that
the $(p,-q)$ torus knot is the mirror image of the $(p,q)$ torus knot
and the $(-p,-q)$ torus knot is equivalent to the $(p,q)$-torus knot
except for the reversed orientation.
In the following, we assume that $p,q>0$ without loss of generality.

\xxx

Suppose that the $(p,q)$-torus knot, designated by $\tT$, is subjected to 
a uniform magnetic field $\vB$.
When an electron migrates between two points $\vecr_a$ and $\vecr_b$ along $\tT$,
it earns an additional quantum phase $\chiab$ defined by
\begin{equation}
\chiab(\vB) = \frac{-e}{\hbar} \int_{\vecr_a}^{\vecr_b} \vA\cdot d\vecr. \quad (e>0)
\label{eq_chi_ini}
\end{equation}
Here, $\vA \equiv (\vB \times \vecr)/2$ is the vector potential associated with $\vB$.
The total phase shift $\chi_{\tT}$ for
the electron going around $\tT$ reads as
\begin{equation}
\chi_{\tT}(\vB) = - 2\pi \frac{\Phi_{\tau}(\vB)}{\Phi_0},
\label{eq_chiB}
\end{equation}
where $\Phi_{\tau}(\vB) \equiv \int_{S_\tau} \vB \cdot d\vn$ is the sum of fluxes that thread
the multiply self-intersecting $p$ surfaces surrounded by ${\cal T}$,
and $\vn$ is the unit vector normal to the surface
whose total area is designated by $S_{\tau}$.

\xxx

We are ready to derive the persistent current $I$ in a $\ell$-length torus knot 
containing $N$ electrons with the effective mass $m^*$.
The current $I_n$ carried by a single electron in the $n$th eigenstate
is $I_n = ev_n/\ell = e \hbar k_n/(m^* \ell)$, where the wavenumber $k_n$ is given by
\begin{equation}
k_n = \frac{2\pi}{\ell} \left( n - \frac{\Phi_{\tau}(\vB)}{\Phi_0} \right),
\quad n=0,\pm1, \pm2, ...
\end{equation}
The total current $I$ at zero temperature
is obtained by summing the contributions from all eigenstates
with energies below the Fermi level. 
Remind that $I$ for odd $N$, denoted by $I_{\rm odd}$, 
differs from that for even $N$, denoted by $I_{\rm even}$,
In fact, we can prove that
\begin{equation}
I_{\rm odd} = 2 \times \sum_{n=-(N-1)/2}^{(N-1)/2} I_n = - I_0 \frac{\Phi_{\tau}(\vB)}{\Phi_0}
\end{equation}
for $-1/2 < \Phi_{\tau}/\Phi_0 < 1/2$, and
\begin{equation}
I_{\rm even} = 2 \times \sum_{n=-N/2+1}^{N/2} I_n = - I_0 \left( \frac{\Phi_{\tau}(\vB)}{\Phi_0} -\frac12 \right)
\end{equation}
for $0 < \Phi_{\tau}/\Phi_0 < 1$
with the definition of $I_0 \equiv 2Neh/(m^* \ell^2)$;
Note that $I_{\rm odd}$ and $I_{\rm even}$
have the same period such that $I_{\alpha}(\Phi_{\tau}) = I_{\alpha}(\Phi_{\tau} + \Phi_0)$
[$\alpha =$ odd or even]
and $I_{\alpha}(0) = 0$.
Since precise control of $N$ is difficult in experiments,
we assume an ensemble average over many 
realizations of quantum torus knots to obtain $(I_{\rm odd} + I_{\rm even})/2$.
As a result, we arrive at
\begin{equation}
I = - I_0 \left( \frac{\Phi_{\tau}(\vB)}{\Phi_0} - \frac14 \right) 
\quad {\rm for} \;\; 0 < \frac{\Phi_{\tau}}{\Phi_0} < \frac12,
\label{eq_007}
\end{equation}
and $I=0$ at $\Phi_{\tau} = 0$, $I(\Phi_{\tau}) = I(\Phi_{\tau}+(\Phi_0/2))$.
It follows from Eq.~(\ref{eq_007}) that every geometric information (except for $\ell$)
is involved in the flux-sum $\Phi_{\tau}$.
The remained task is, therefore, to reveal the dependencies of $\Phi_{\tau}(\vB)$ on
the direction of the field $\vB$ and the geometry of the knot.

\xxx

To deduce the explicit form of $\Phi_{\tau}$ (or equivalently $\chi_{\tT}$), 
we use the relation
\begin{equation}
d\vecr = \frac{d\vecr}{d\te} d\te
= \left( \frac{d u_i}{d\te} \bm{e}_i \right) d\te
\;\; \mbox{and} \;\;
\vB = B_i \bm{e}_i,
\label{eq_drdte}
\end{equation}
where $u_i = x,y,z$ and $B_i = B_x, B_y, B_z$ for $i=1,2,3$, respectively,
and $\bm{e}_i$ is the $u_i$-directed unit vector \cite{Higher};
the summation convention with respect to $i$ was used in Eq.~(\ref{eq_drdte}).
From Eqs.~(\ref{eq_chi_ini}) and (\ref{eq_drdte}), we find
\begin{eqnarray}
\chi_{\tT}(\vB)
&=&
\frac{-e}{2\hbar}
\int_0^{2\pi} 
\Big[
B_z \left( x y' - y x' \right) \nonumber \\
& &+
B_x \left( y z' - z y' \right)
+
B_y \left( z x' - x z' \right)
\Big] d\te,
\label{eq_chi_new}
\end{eqnarray}
where $x' \equiv dx/d\theta$ so do $y'$ and $z'$.
Referring to Eq.~(\ref{eq_xyz}),
the integrand of Eq.~(\ref{eq_chi_new}) is written in terms of a trigonometric series.
For instance, we have
\begin{equation}
x y' - y x'
=
p R^2 + 2p R\ep \cos(q\te) + \frac{p}{2} \ep^2 \Big[ 1+\cos(2q\te) \Big],
\label{eq_reduce}
\end{equation}
and thus straightforward integration yields
\begin{equation}
\int_0^{2\pi} \left( x y' - y x' \right) d\te
=
2p \left( \pi R^2 + \frac{\pi}{2} \ep^2 \right),
\label{eq_integ_proof}
\end{equation}
where we took into account that $p(\ne 0)$ and $q(\ne 0)$ are relatively prime integers.

\begin{figure}[ttt]
\centerline{
\includegraphics[width=7.5cm]{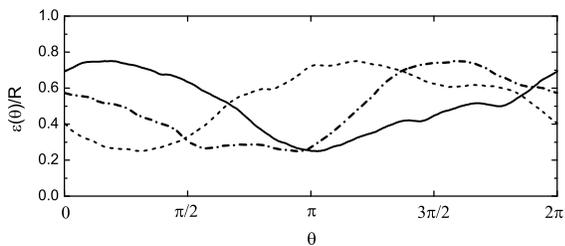}
}
\caption{Landscape of spatially modulated minor radius $\ep(\theta)$
of the torus knot.
Three typical patterns are presented,
all of which fluctuate across the centerline $\bar{\ep}=R/2$ with the amplitude $\delta \ep = R/4$.
}
\label{fig_2}
\end{figure}

In a similar manner, it is easy to derive
\begin{eqnarray}
& &
y z' - z y' \nonumber \\ [4pt]
&=&
\frac{R\ep}{2}
\Big\{ -(p-q) \sin[(p+q)\te] + (p+q) \sin[(p-q)\te] \Big\} \nonumber \\
&+& \!
\frac{\ep^2}{4} \!
\Big\{
4q\sin(p\te) \! + \! p\sin[(p \!-\! 2q)\te] \! - \! p \sin[(p\!+\!2q)\te]
\Big\},
\end{eqnarray}
and
\begin{eqnarray}
& &
z x' - x z' \nonumber \\ [4pt]
&=&
\frac{R\ep}{2}
\Big\{ (p-q) \cos[(p+q)\te] - (p+q) \cos[(p-q)\te] \Big\} \nonumber \\
&-& \!
\frac{\ep^2}{4} \!
\Big\{
4q\cos(p\te) \! + \! p\cos[(p\!-\!2q)\te] \! - \! p \cos[(p\!+\!2q)\te]
\Big\}.
\end{eqnarray}
Note that $p\pm q \ne 0$ and $p \pm 2q \ne 0$, since $p,q$ are coprime.
Hence we have
\begin{equation}
\int_0^{2\pi} \left( y z' - z y' \right) d\te = \int_0^{2\pi} \left( z x' - x z' \right) d\te = 0.
\label{eq_integ_proof2}
\end{equation}
Substituting the results into Eq.~(\ref{eq_chi_new}), we obtain
\begin{equation}
\chi_{\tT}(\vB)
=
\frac{-e}{\hbar} \times p B_z \left( \pi R^2 + \frac{\pi}{2}\ep^2 \right),
\label{eq_chiT_new}
\end{equation}
and
\begin{equation}
\Phi_{\tau}(\vB) = p B_z \left( \pi R^2 + \frac{\pi}{2}\ep^2 \right).
\label{eq_PhiSt}
\end{equation}

\xxx

The formula (\ref{eq_PhiSt}) is the main fruit of this article, describing the effects
of the knotted geometry and field direction
on the periodic oscillation of $I(\Phi_{\tau})$.
The following three peculiarities can be drawn.
First, the transverse components of the field, $B_x$ and $B_y$,
play no role in the quantum interference in the present systems.
{\it Only} the vertical component $B_z$ determines both the quantum phase shift $\chi_{\tau}$
and the oscillation period of $I$, although the other two components should be
reckoned in the contour integral defined by Eq.(\ref{eq_chi_ini}).
Vanishing the $B_x$- and $B_y$-contributions owes to the zero-sum rule
expressed by Eq.~(\ref{eq_integ_proof2}),
which holds for any choice of $(p,q)$.

\xxx

Second,
$I(\Phi_{\tau})$ in $(p,q)$-torus knots is independent of $q$,
depending {\it only} on $p$ in the form of $I \propto p$.
This fact is trivial if $\ep \to 0$, namely, the knot is reduced to a simple $R$-radius circle
along which the electron travels $p$ times for one duration.
But for finite $\ep$, $(p,q)$-torus knots deviate geometrically from
the $R$-radius circle and those with different values of $q$ (but the same value of $p$)
exhibit different entangled structures extending over the three-dimensional space,
which makes the $q$-independence unobvious.

\begin{figure}[ttt]
\centerline{
\includegraphics[width=7.2cm]{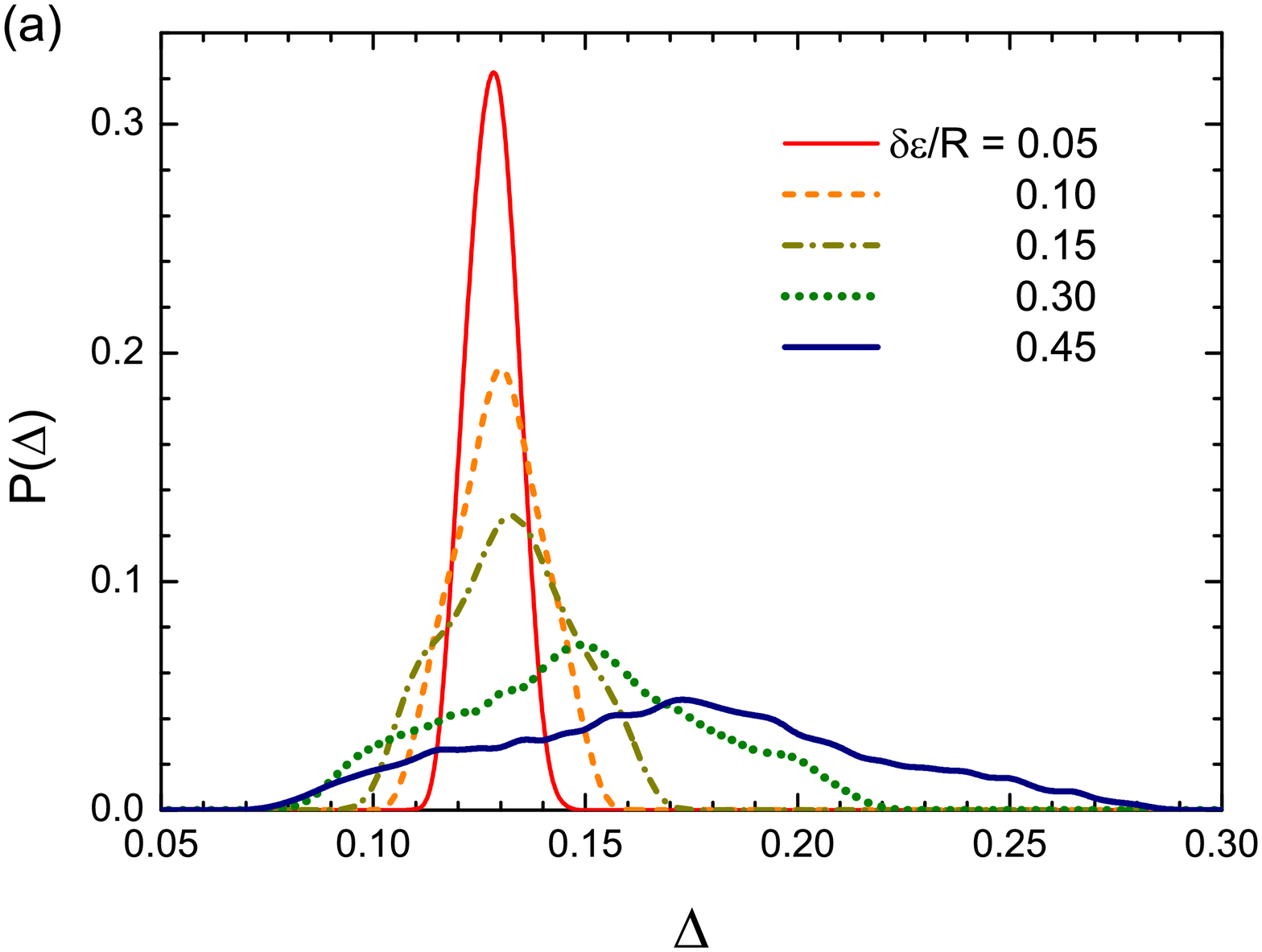}
}
\centerline{
\includegraphics[width=7.2cm]{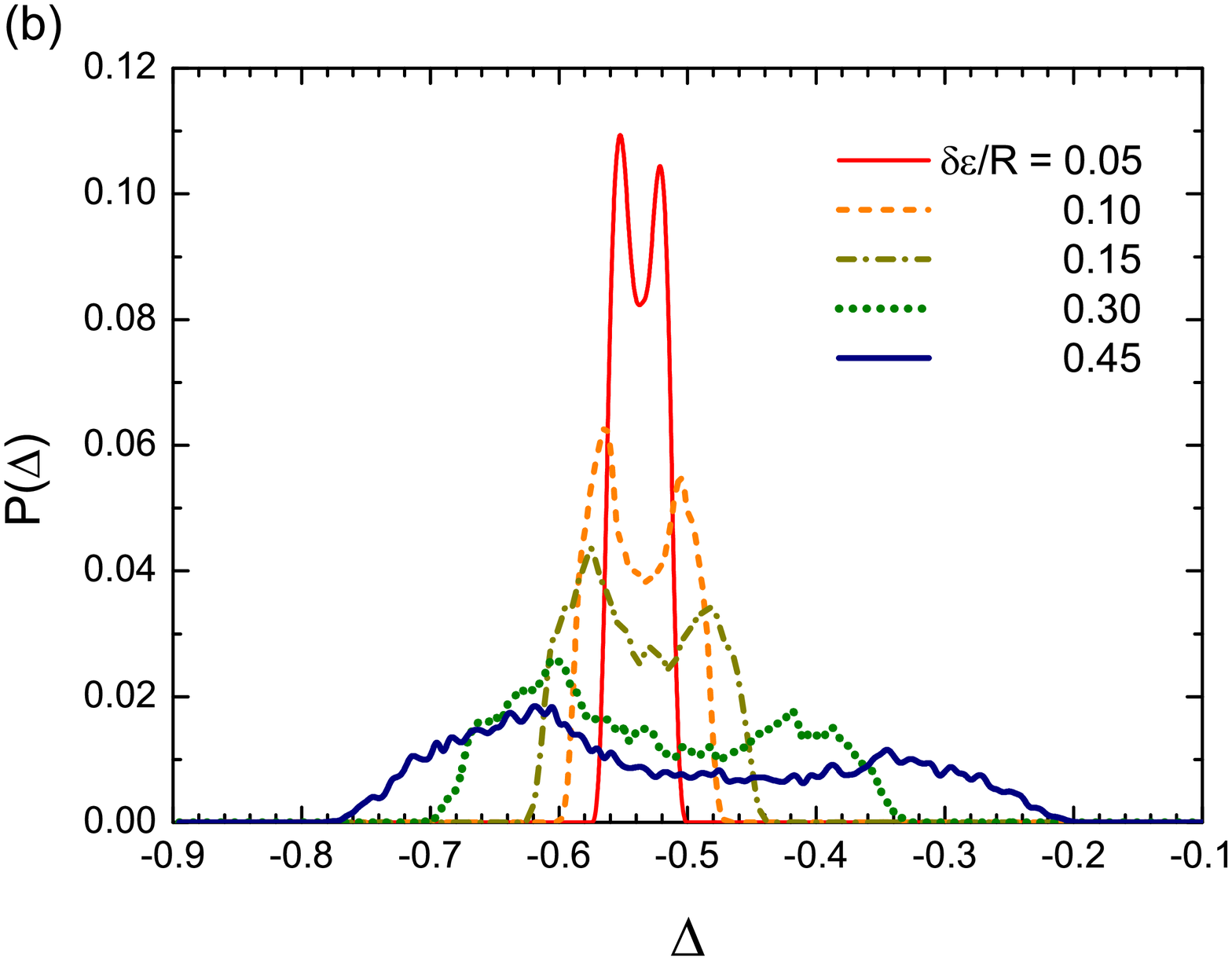}
}
\caption{(color online) Probability distribution function $P(\Delta)$ of the excess term $\Delta$
defined by Eq.~(\ref{eq_excess}).
The field direction is set as
$\vB/|\vB| = (0,0,1)$ for (a) and $\vB/|\vB| = (\frac{1}{\sqrt{2}},\frac{1}{\sqrt{3}},\frac{1}{\sqrt{6}})$ for (b).}
\label{fig_3}
\end{figure}

\xxx

Third, the minor radius $\ep$ gives to $\Phi_{\tau}$ an excess contribution
on the order of $\ep^2$.
For real knotted materials, it may amount to $\ep \sim R/2$ more or less;
hence the excess term $\pi\ep^2/2$ shown in Eq.~(\ref{eq_PhiSt}) 
causes $\Phi_{\tau}$ to increase about ten percent compared with the case of $\ep=0$,
no matter what value $p$ and $q$ possess.
It should be emphasized that all the three peculiarities listed above are implications
of the single formula (\ref{eq_PhiSt}),
which we have analytically derived with neither aid of numerical computation
nor mathematical approximation.

\xxx

One may be interested in whether geometric distortion of the knots
causes a change in the formula (\ref{eq_PhiSt}).
It is certain that the trajectories of actual knotted materials deviate from the ideal torus knot geometry
because of mechanical distortion and thermal disturbance.
Such the imperfection effects can be evaluated in part by considering a distorted torus knot
whose minor radius $\ep$ is spatially modulated,
as demonstrated below.

\xxx

Figure \ref{fig_2} shows typical profiles of the $\te$-dependent $\ep(\te)$
we have considered, wherein every curves fluctuate smoothly around the centerline $\bar{\ep}=R/2$
with the amplitude of $\delta \ep = R/4$.
These modulated $\ep(\theta)$ are generated by the Fourier filtering method; 
see Refs.~\cite{PrakashPRA1992,ShimaPRB2004} for details.
The spatial modulation in $\ep$ leads a breakdown of the zero-sum rules (\ref{eq_integ_proof2}),
thus revives the $B_x$- and $B_y$-contributions in the integrand of Eq.~(\ref{eq_chi_new}).
As a result, the formula (\ref{eq_PhiSt}) should be revised as
\begin{equation}
\Phi_{\tau}(\vB) = p B_z \times \pi R^2 \left[ \;1 + \;\Delta (\vB,\delta \ep)\; \right].
\label{eq_excess}
\end{equation}
The excess term $\Delta$ takes various values depending on the $\ep(\theta)$ configuration
of individual distorted $(p,q)$ torus knots.
Hence, we evaluate $\Delta$ for various $\ep(\te)$-realizations and field directions
to obtain the probability with which a specific value of $\Delta$ occurs.

\xxx

Figure \ref{fig_3} demonstrates the probability distribution functions $P(\Delta)$
under the conditions of:
$\vB/|\vB| = (0,0,1)$ for (a) and $\vB/|\vB| = (\frac{1}{\sqrt{2}},\frac{1}{\sqrt{3}},\frac{1}{\sqrt{6}})$ for (b).
We set $\bar{\ep}=R/2$ and $(p,q)=(2,3)$ for all the plots.
It is found in Fig.~\ref{fig_3}(a) that a sharp peak emerging at $\delta \ep \ll R$
becomes broadened with increasing $\delta \ep$,
together with the slight shift in the peak position from $\Delta \sim 0.13$ to $\Delta \sim 0.17$.
Similar behavior is observed in Fig.~\ref{fig_3}(b),
while two peaks (instead of one) moving apart from each other are observed.
Attention should be paid for that in the latter panel,
$\Delta$ for large $\delta \ep$ reaches minus one in rare setting,
which results in $\Phi_{\tau} \equiv 0$ as seen from Eq.~(\ref{eq_excess}).
The same situation takes place incidentally when the field direction
deviates significantly from the vertical one.
This fact implies the preference for distortion-free knots preparation
in order to observe a well-defined periodic oscillation of $I(\Phi_{\tau})$in experiments.

\xxx

As a closing remark, it is worthy to mention that the history of persistent currents
dates back to the early days of quantum mechanics;
nonzero orbital angular momentum of aromatic molecules
were already noticed many decades ago \cite{PaulingJCP1936,LonsdalePRSLA1937}.
Nowadays, it is revealed that
realistic knotted materials may have twisted internal structures
such as twisted $\pi$-electron orbitals in torus-knot-shaped organic molecules \cite{WannereJPhysChemA2009}.
The twisted nature is known to affect the aromaticity of knotted molecules \cite{Miliordos}
or yield zero-field quantum phase shift \cite{Taira},
which we have omitted in this work.
It is also interesting to note that quantum confinement into bend hollow cylinders
such as deformed carbon nanotubes \cite{VilatelaAdvMater2010,ShimaCNT}
produce an effective electro-static potential field;
the geometry-induced potential can result in a qualitative change in single-particle states
\cite{AtanasovPLA2009,JensenPLA2011,AtanasovJPA2012} 
and collective excitations \cite{ShimaPRB2009,OnoEPL2011} of electrons
in the system.
Though the present work is based on a simplified model,
I believe that the results obtained provoke further quantitative discussions
as to knotted material properties taking into account detailed atomic/molecular configuration
and the geometry-induced field effects commented above.

\xxx

This study was supported by a Grant-in-Aid for Scientific Research from the MEXT, Japan.
The author acknowledges K.~Yakubo, Y.~Asano, S.~Ono, K.~Izumi for valuable comments
and M.~Sato for his analytic support.



\end{document}